\pdfoutput=1
\documentclass[table]{ccjnl}
\usepackage{lipsum,amsmath}
\usepackage{cuted}
\usepackage{bm}
\usepackage{tabularx} 
\graphicspath{{figures/}}
\usepackage{booktabs}

\usepackage{multirow}
\newcolumntype{Y}{>{\centering\arraybackslash}X}
\title{AI for CSI Prediction in 5G-Advanced and Beyond}
\author{Chengyong Jiang\inst{1}, Jiajia Guo\inst{1,2}, Xiangyi Li\inst{1}, Shi Jin\inst{1,*}\corinfo{jinshi@seu.edu.cn}, Jun Zhang\inst{2}}
\receiveddate{xxx. xx, 2024}
\reviseddate{xxx. xx, 2024}
\Editor{xxx xxx}

\address[1]{National Mobile Communications Research Laboratory, Southeast University, Nanjing
	210096, China}
\address[2]{The Department of Electronic and Computer Engineering, Hong Kong University of Science and Technology, Hong Kong 999077, China}

\begin{document}

\maketitle

\begin{abstract}
Artificial intelligence (AI) is pivotal in advancing fifth-generation (5G)-Advanced and sixth-generation systems, capturing substantial research interest. Both the 3rd Generation Partnership Project (3GPP) and leading corporations champion AI's standardization in wireless communication. This piece delves into AI's role in channel state information (CSI) prediction, a sub-use case acknowledged in 5G-Advanced by the 3GPP. We offer an exhaustive survey of AI-driven CSI prediction, highlighting crucial elements like accuracy, generalization, and complexity. Further, we touch on the practical side of model management, encompassing training, monitoring, and data gathering. Moreover, we explore prospects for CSI prediction in future wireless communication systems, entailing integrated design with feedback, multitasking synergy, and predictions in rapid scenarios. This article seeks to be a touchstone for subsequent research in this burgeoning domain.
\keywords{Massive MIMO; CSI prediction; Artificial intelligence; 5G-Advanced; 3GPP; Standardization}
\end{abstract}
\section{Introduction}
\label{s1}
Artificial intelligence (AI) has found significant success in diverse areas of wireless communications, owing to its rapid technological advancements in recent years \cite{8233654,10158439,wang2023road}. In stark contrast to traditional methods, AI-based approaches possess the remarkable capability to automatically and directly extract features from vast training datasets via neural networks (NNs), thereby eliminating the necessity for explicit domain-specific knowledge analysis. The fusion of AI into wireless communication systems has emerged as a pivotal trend in the evolution towards fifth-generation (5G)-Advanced and sixth-generation networks \cite{imt,10121037}, eliciting substantial research endeavors from both academia and industry. This technology has enabled its seamless integration and effective deployment across multiple domains within wireless communications, including channel feedback \cite{8322184,lu2018mimo,10495862}, beam management \cite{8542687,8968715}, and localization \cite{9390409,10163845}.

Standardization efforts for AI in wireless communication are actively being pursued by the 3rd Generation Partnership Project (3GPP) and various corporations. The study item ``Study on Artificial Intelligence (AI)/Machine Learning (ML) for NR Air Interface'' has been approved by the 3GPP, with a focus on exploring the benefits of AI-based algorithms and assessing their standardization impacts on existing air interfaces. Representative use cases, such as channel state information (CSI) feedback enhancement, beam management, and positioning accuracy enhancement, have been selected to establish the groundwork for future AI research in the air interface \cite{213599}. In November 2022, the 3GPP meeting endorsed two initial sub-use cases for ``CSI feedback enhancement'', namely CSI compression using a two-side model \cite{guo2022ai,9931713} and CSI prediction using a user equipment (UE)-side model.

This article specifically focuses on the sub-use case of AI for CSI prediction. The implementation of the massive multiple-input multiple-output (MIMO) technology, with numerous transmit antennas equipped at the 5G New Radio (NR) NodeB (gNB), significantly enhances spectrum and energy efficiency. However, the benefits of massive MIMO heavily rely on accurate CSI at the gNB \cite{love2008overview}. Improving the accuracy of CSI is crucial for optimizing communication performance. However, the motions of user equipment (UE) introduce a delay between CSI estimation and utilization, known as the \textit{channel aging problem}. Channel aging impedes the acquisition of accurate CSI, especially in high-speed scenarios, leading to degraded MIMO performance \cite{7473866}. Furthermore, CSI estimation and feedback can lead to excessive overhead and strain precious bandwidth due to the increasing number of antennas, making overhead reduction a challenging issue for massive MIMO.

In response to these challenges, CSI prediction has emerged as a promising approach to forecasting future CSI based on historical data. 
Recent advancements have explored the potential of AI in CSI prediction.
AI-based CSI prediction methodologies treat the dynamic channel as a time series, framing the prediction task as a regression problem. NNs are employed to capture and leverage channel characteristics from abundant channel data, eliminating the need for prior channel knowledge. This approach yields substantial performance gains over conventional methods, underscoring the potential of AI in enhancing CSI prediction capabilities.
AI was initially introduced for CSI prediction in \cite{8395053}, where the OCEAN prediction framework was presented. This framework not only demonstrated high accuracy in CSI prediction but also facilitated rapid acquisition of the predicted CSI.
Subsequent advancements included a transformer-based parallel CSI prediction network with a pilot-to-precoder prediction scheme, as described in \cite{pre3}. Additionally, a reinforcement learning-driven network for joint CSI prediction and beamforming in multi-user scenarios was proposed in \cite{pre4}. To further enhance efficiency, transfer learning was integrated to reduce training times and data costs, as evidenced in \cite{10174691} and \cite{9175003}.
CSI prediction in unique communication scenarios such as low earth orbit satellite and high-speed railway were explored in \cite{9439942} and \cite{hsr}, respectively.
Moreover, \cite{8979256} introduced an AI-autoregressive (AR) predictor hybrid for time-division duplex scenarios. To dynamically adapt to wireless channel variations, a hypernetwork-based CSI prediction framework was proposed in \cite{10404045}, enabling the dynamic updating of NN parameters.
Addressing the high computational complexity associated with AI-based methods, a low-complexity CSI prediction approach was devised in \cite{10333396}. The validity and practicality of AI-based CSI prediction methods were affirmed through real-world data experiments reported in \cite{9691478}.

Although AI-based CSI prediction methods have achieved significant performance gains in current studies, the standardization of AI for CSI prediction still faces several unresolved issues, necessitating further exploration and discussion. For instance, most of the existing works are based on simulations, whose assumptions may diverge from the realities of actual systems. Moreover, model management operations such as model training, model monitoring, and data collection are vital to the practical deployment of AI, known as lifecycle management problems. Model management is usually not considered exactly in the existing works, which requires clarification and regulation for further standardization.

This article provides an in-depth exploration of AI in CSI prediction for 5G-Advanced and future wireless systems. For insights of AI in CSI compression, please refer to the earlier study \cite{guo2022ai}. Unlike previous surveys on AI for CSI prediction \cite{82336541,82336542,82336543}, this article does not concentrate on specific prediction frameworks or NN structures. Instead, it delves into a unique standardization topic, incorporating extensive discussions from the 3GPP. Specifically, it offers a detailed comparison of various prediction methodologies, addresses the problems of model management, and anticipates AI-based predictions in future wireless communication systems, drawing insights from Rel-19 discussions \cite{234039,sum116} and the International Telecommunication Union's (ITU) vision for IMT-2030 \cite{imt}.
The subsequent sections are organized as follows: Section \ref{s2} introduces and compares representative non-AI and AI-based prediction methods. In Section \ref{s3}, the scope of AI-based prediction is defined at first. The performance evaluation methodology for AI-based prediction methods, encompassing prediction accuracy, model generalization, and model complexity evaluation, is then discussed. Furthermore, model management issues for AI-based prediction, including model training, model monitoring, and data collection, are addressed. Section \ref{s4} introduces prospects for AI for CSI prediction, involving integrated designs with CSI feedback, multi-tasking integration, and predictions for high-speed situations.
Section \ref{s5} concludes this work. The primary literature reviewed for the standardization of AI for CSI prediction is summarized in Table \ref{Category}.

\begin{table*}[htbp]
\centering
\caption{Summary of the primary literature reviewed for the standardization of AI for CSI prediction.}
\begin{tabular}{l l l}
\toprule \label{Category}
\textbf{Category} & \textbf{Ref.} & \textbf{Contribution} \\
\midrule
Background and existing papers  & \cite{10345638,9376324,9994050,10411052,9076084,10279462} & Provide background information on various standardization aspects and vision. \\ 
\midrule
\makecell {Standardization objectives and vision}  & \cite{imt} & Present the vision for the next generation of wireless communication. \\ & \cite{213599} & Establish standardization study objectives of the use case ``CSI feedback enhancement''. \\
 & \cite{234039} & Propose specific working objectives for AI for CSI prediction in R19. \\
\midrule
Evaluation results & \cite{oppo,9210016} & Offer a performance comparison between non-AI and AI-based prediction methods. \\
 & \cite{zte} & Compare channel matrix and eigenvector prediction methods in terms of effectiveness. \\
 & \cite{cell,10262359} & Compare the performance of cell-specific, cell-common, and no-prediction methods. \\
 & \cite{9175003,2400795} & Verify the effectiveness of transfer learning in AI-based CSI prediction. \\
 & \cite{speed} & Evaluate the generalization ability of prediction methods across different speeds. \\
\midrule
Standardization insights   & \cite{guo2022ai} & Discuss the combination of AI-based CSI compression and prediction. \\ & \cite{sum116} & Provide insights on various standardization aspects for AI-based CSI prediction. \\ & \cite{2402267} & Propose a gNB-side training method with model delivery to train the UE-side model. \\
 & \cite{pse} & Introduce a power spectral entropy-based model monitoring method. \\
 & \cite{att} & Propose to specify a new data collection framework for model management requirements. \\
 & \cite{2402026} & Discuss the considerations of data collection for different model management purposes. \\

\bottomrule
\end{tabular}
\end{table*}

\section{Non-AI Vs. AI-based CSI Prediction}
\label{s2}
In this section, we introduce two representative non-AI CSI prediction methods followed by three AI-based CSI prediction methods. A comparison of these methods is then presented.

\subsection{Non-AI CSI Prediction Methods}

Existing non-AI CSI prediction methods can be broadly categorized into two types: AR and parametric model-based methods.

1) \textit{AR Model-based Methods}:
\label{ar} AR model-based CSI prediction methods model the fading channel as an autoregressive process, enabling the extrapolation of future CSI through a combination of historical CSI using a Wiener filter or Kalman filter to estimate AR coefficients \cite{9210016}. While AR model-based methods perform well under slow-changing channels, they require continuous updates of model parameters to adapt to changing environments. This leads to significant computational complexity, limiting their performance in fast-changing channels. Moreover, the complexity of AR model-based methods increases significantly with the number of antennas, and a long channel observation time is required for precise prediction, making them less practical for massive MIMO systems.

2) \textit{Parametric Model-based Methods}:
\label{sos}Parametric model-based methods represent the fading channel as a superposition of complex sinusoidal signals. These methods assume that the model parameters, such as complex path gain, angles of arrival and departure, Doppler Shift, and the number of multipath, remain constant or vary slowly during the prediction time. The future CSI can be predicted based on these estimated parameters from historical CSI \cite{sos}. However, the parameter estimation process of parametric model-based methods is complex and time-consuming. Additionally, model parameters may become outdated over time, especially in high-speed scenarios, resulting in degraded performance. Frequent updates of model parameters are required to adapt to changes in the wireless channel, which leads to high complexity.

\subsection{AI-based CSI Prediction Methods}
\begin{figure*}[t]
  \centering
  \includegraphics[width=0.8 \textwidth]{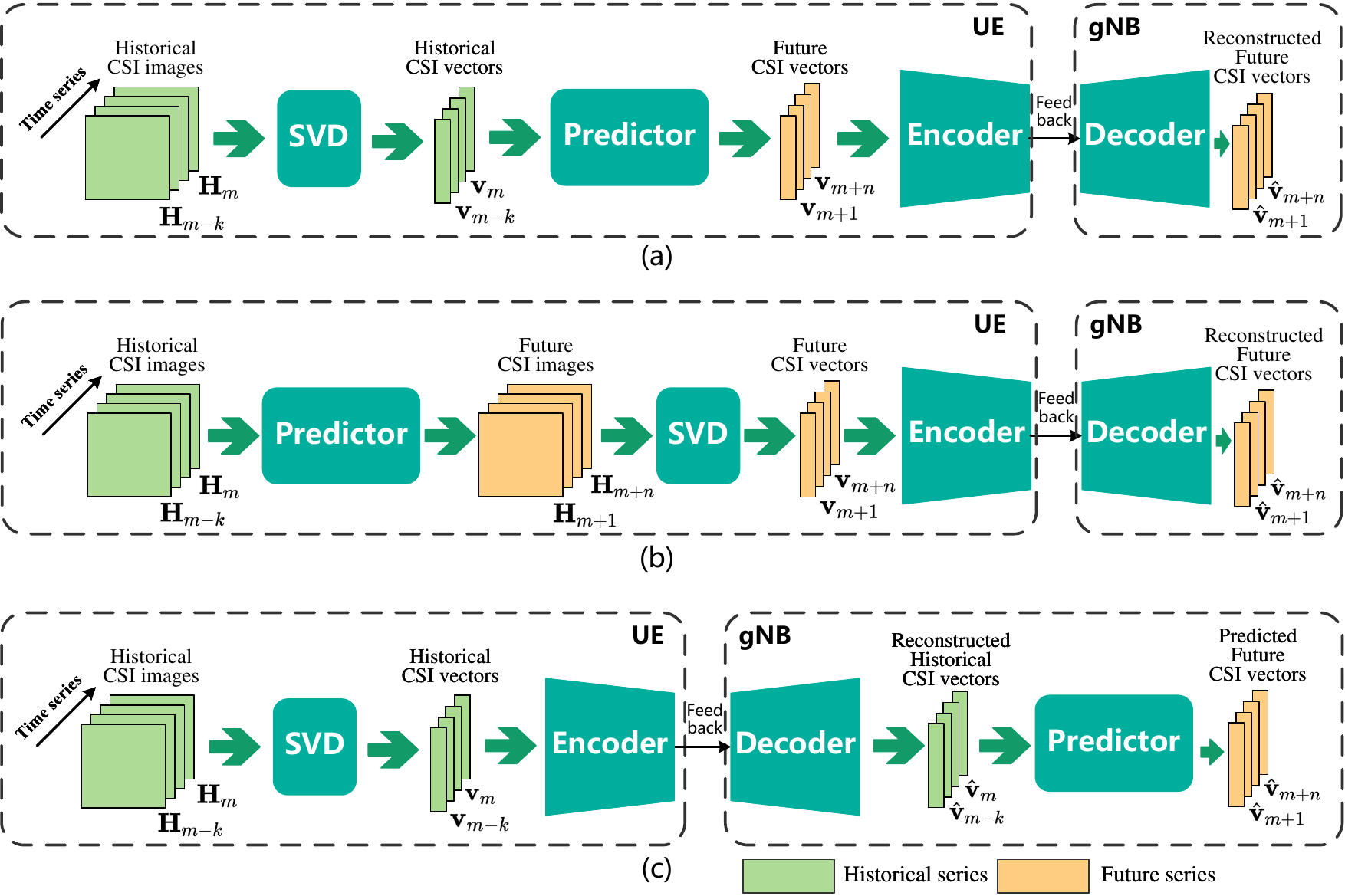}
  \caption{Frameworks of AI-based CSI prediction methods, including: (a) Eigenvector prediction using a UE-side model; (b) Channel matrix prediction using a UE-side model; (c) Eigenvector prediction using a gNB-side model.}
  \label{pm}
  \vspace{-2mm}
\end{figure*}

As per the current discussion in the 3GPP, AI-based CSI prediction methods can be discerned based on the predictor's location and the type of input and output data. Thus, existing AI-based CSI prediction frameworks can be divided into three types: eigenvector prediction using a UE-side model, channel matrix prediction using a UE-side model, and eigenvector prediction using a gNB-side model, as shown in Fig. \ref{pm}. 

1) \textit{Eigenvector Prediction Using a UE-side Model}:
\label{}In this framework (Fig. \ref{pm}(a)), the predictor is located at the UE side, and both the input and output of the predictor are eigenvectors. The method involves performing singular value decomposition (SVD) on the complete channel matrix $\mathbf{H}$ to obtain eigenvectors $\mathbf{v}$. A series of historical eigenvectors $\left\{\mathbf{v}_{m-k}, \ldots, \mathbf{v}_{m}\right\}$ are then fed into the predictor to predict future eigenvectors $\left\{\mathbf{v}_{m+1}, \ldots, \mathbf{v}_{m+n}\right\}$. The predicted eigenvectors are compressed and sent back to the gNB, which reconstructs the eigenvectors based on the received feedback information.

2) \textit{Channel Matrix Prediction Using a UE-side Model}:
Fig. \ref{pm}(b) illustrates the UE-side channel matrix prediction framework, which is widely adopted in current academic studies \cite{guo2022ai}. Here, the UE predicts future channel matrices $\left\{\mathbf{H}_{m+1}, \ldots, \mathbf{H}_{m+n}\right\}$ based on historical channel matrices $\left\{\mathbf{H}_{m-k}, \ldots, \mathbf{H}_{m}\right\}$, followed by SVD to obtain corresponding eigenvectors $\left\{\mathbf{v}_{m+1}, \ldots, \mathbf{v}_{m+n}\right\}$. The UE then compresses the eigenvectors for feedback, and the gNB reconstructs the original eigenvectors based on the received feedback information.

3) \textit{Eigenvector Prediction Using a gNB-side Model}
Fig. \ref{pm}(c) represents the gNB-side eigenvector prediction method. The predictor is situated at the gNB side, and both the input and output of the predictor are eigenvectors due to the implicit feedback mechanism. In this approach, the UE first sends back $\left\{\mathbf{v}_{m-k}, \ldots, \mathbf{v}_{m}\right\}$ to the gNB. The gNB then predicts future eigenvectors $\left\{\mathbf{\hat{v}}_{m+1}, \ldots, \mathbf{\hat{v}}_{m+n}\right\}$ based on the previously received and reconstructed eigenvectors series $\left\{\mathbf{\hat{v}}_{m-k}, \ldots, \mathbf{\hat{v}}_{m}\right\}$, eliminating the need for CSI feedback for those time steps. Consequently, this framework reduces the additional feedback overhead compared to the first two frameworks.

\subsection{Comparison}
1) \textit{Non-AI Vs. AI-based Prediction Methods}:
\label{Non-AI}
\begin{figure}[t]
  \centering
  \includegraphics[width=0.40\textwidth]{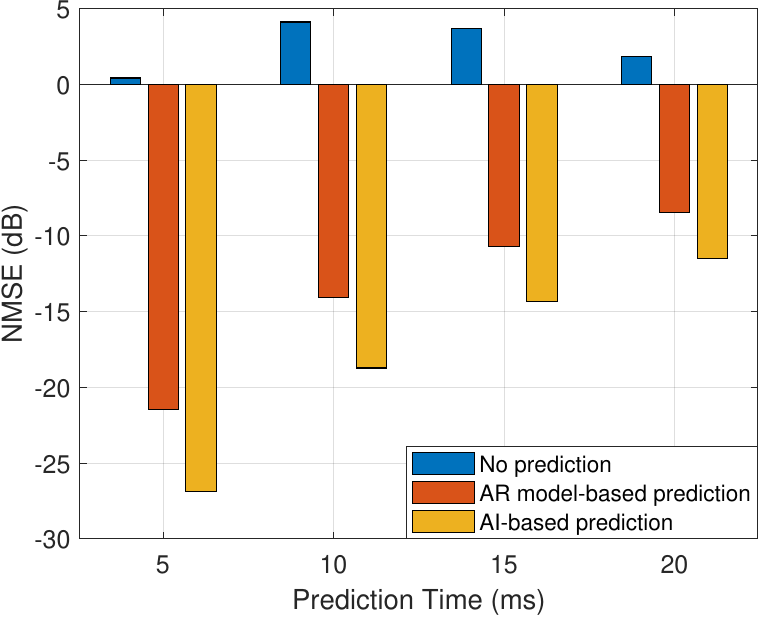}
  \caption{Comparison of AI-based, non-AI, and no prediction methods \cite{oppo}. }
  \label{non}
  \vspace{-2mm}
\end{figure}
Fig. \ref{non} displays a performance comparison between non-AI (i.e., AR model-based) and AI-based prediction methods \cite{oppo} for channel matrices. 
The channel configuration comprises 32 transmit antennas and 4 receive antennas. The CSI-RS is configured with a periodicity of 5ms. Simulations are performed in the outdoor scenario with 30km/h UE speed. The no prediction method serves as the baseline, comparing the current CSI with the latest historical CSI for evaluation. Normalized mean-squared error (NMSE) is used as the evaluation metric. 
It is noteworthy that the issue of intellectual property protection has led to discrepancies in the underlying simulation configurations and assumptions among different entities. This includes variations in model architectures, training processes, and data generation settings, which result in differing outcomes. Unfortunately, some of these detailed aspects are not fully disclosed in the source documents. Consequently, only results generated under a unified configuration can be considered  comparable. For those seeking further clarification and in-depth discussions, it is recommended to consult the original source document.

The results in Fig. \ref{non} indicate that AI-based prediction significantly outperforms AR model-based and no prediction methods. Furthermore, AI-based prediction allows for more flexible settings of prediction intervals, making it suitable for practical use. However, current comparisons between non-AI and AI-based prediction methods mainly focus on prediction accuracy and computational complexity, neglecting the model management-related overhead of AI-based methods. Further research is required for a more comprehensive and fair comparison. Further research is required for a more comprehensive and fair comparison.

2) \textit{Eigenvector Vs. Channel Matrix Prediction Methods}:
\begin{table}[t]  
\begin{minipage}{\columnwidth}   
\centering  
\footnotesize
\caption{Comparison of channel matrix and eigenvector prediction methods in SGCS \cite{zte}.}  
\label{ce}  
\begin{tabularx}{\columnwidth}{Ycc}   
\toprule  
\addlinespace[0.5em]  
\multirow{2}{*}{Prediction Time} & \multicolumn{2}{c}{Prediction Method} \\  
\cmidrule(lr){2-3}  
\addlinespace[0.2em] 
& Channel Matrix & Eigenvector \\  
\midrule  
5\,ms & \cellcolor[HTML]{EFEFEF}\hspace{0.5em}0.989\hspace{0.5em} & \hspace{0.5em}0.844\hspace{0.5em} \\  
10\,ms & \cellcolor[HTML]{EFEFEF}\hspace{0.5em}0.906\hspace{0.5em} & \hspace{0.5em}0.768\hspace{0.5em} \\  
15\,ms & \cellcolor[HTML]{EFEFEF}\hspace{0.5em}0.806\hspace{0.5em} & \hspace{0.5em}0.746\hspace{0.5em} \\  
\addlinespace[0.2em]   
\bottomrule  
\addlinespace[0.5em] 
\end{tabularx}  
\end{minipage}  
\end{table}  
Table \ref{ce} presents a performance comparison between channel matrix and eigenvector prediction methods using AI in squared generalized cosine similarity (SGCS). 
The ResNet architecture serves as the backbone structure for the predictors. The model inputs consist of 10 historical CSI matrices, which are utilized to predict 3 future CSI matrices. All predictors undergo training on a dataset containing 80,000 samples.
The results indicate that channel matrix prediction, performing prediction before SVD, outperforms eigenvector prediction, which performs prediction after SVD. The superiority of the former can be attributed to the complete channel matrix containing more information, enabling the exploitation of additional channel characteristics for prediction.

However, the channel matrix prediction method has two main challenges. Firstly, it has generally higher complexity compared with the eigenvector prediction method, as channel matrix has larger size. Secondly, the prediction accuracy of the complete channel matrix, evaluated using NMSE, may not match the SGCS performance, leading to discrepancies in propagation scenarios where the eigenvector can maintain high SGCS performance despite degraded NMSE performance.

3) \textit{UE-side Vs. gNB-side Prediction Methods}: \label{gNB}
The gNB-side CSI prediction method involves compressing and reconstructing eigenvectors, which inherently incurs information loss compared to the UE-side prediction framework. To illustrate the effect of CSI compression on CSI prediction accuracy, we conducted a simulation using a CSI predictor comprising four Dense layers. For the CSI compression component, we employed the bi-ImCsiNet architecture proposed in \cite{chen2021deep}. We analyzed the predictor's performance using compression ratios of 1/4, 1/8, 1/16, and 1/32. 
\begin{figure}[t]
  \centering
  \includegraphics[width=0.40\textwidth]{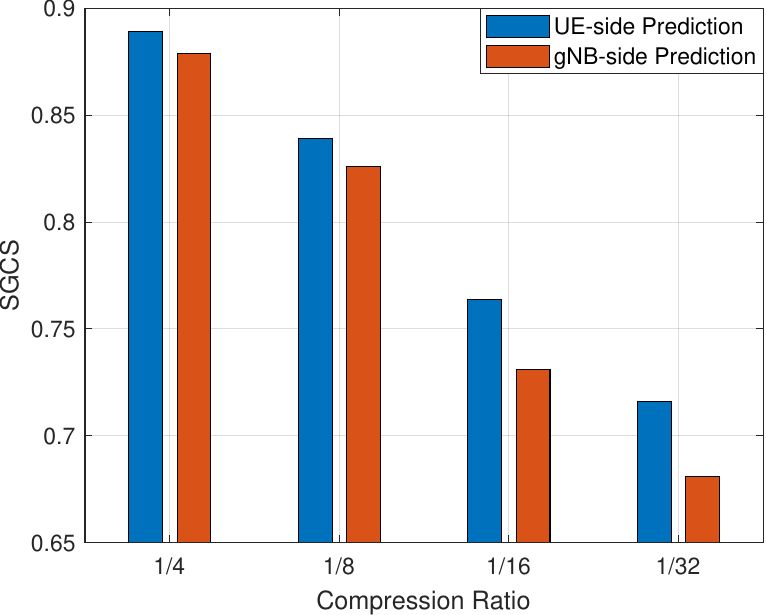}
  \caption{Comparison of UE-side and gNB-side prediction with different compression ratios. }
  \label{gnb}
  \vspace{-2mm}
\end{figure}

The simulation results, depicted in Figure \ref{gnb}, exhibit a distinct pattern: as the compression ratio escalates, the accuracy of the CSI predictions at the gNB diminishes. This decline is attributed to the information loss inherent in the compression process, which subsequently leads to reduced prediction accuracy for the gNB-side prediction.
However, practical 5G systems often face network bottlenecks on the UE side due to limited storage space and computing capacity. Consequently, the complexity of the NN deployed at the UE side is restricted by the UE's capabilities, which may limit prediction performance. Moreover, the gNB-side prediction method can save additional feedback overhead compared to the UE-side prediction method. While CSI prediction using a UE-side model is currently adopted as a sub-use case, the gNB-side prediction is expected to be added as a sub-use case after sufficient study of the UE-side prediction framework.

\section{Key Aspects and Open Problems of AI for CSI Prediction}
\label{s3}

In this section, we elaborate on key aspects of standardizing AI for CSI prediction and discuss open problems pertaining to AI-based prediction methods. We first present the scope of AI-based prediction, including delay compensation, redundant CSI-reference signal (CSI-RS) elimination, and feedback overhead reduction. Subsequently, we introduce the key performance indicators (KPIs) for evaluating AI-based CSI prediction methods, including prediction accuracy, generalization, and complexity evaluation. Lastly, we address model management challenges, including training, monitoring, and data collection.

\subsection{Scope of AI-based CSI Prediction}

The current scope of AI-based CSI prediction encompasses three primary facets: delay compensation, redundant CSI-RS elimination, and feedback overhead reduction.

\subsubsection{Delay Compensation}
In practical systems, there is a delay between estimating and utilizing CSI due to propagation and processing time, known as the channel aging problem. This delay negatively affects the performance of MIMO systems, especially in high-speed scenarios. Channel prediction can be used to effectively address this issue. However, measuring the precise delay in real deployments is challenging. Therefore, the prediction model should adapt to different delays, and it should be capable of tracking the changes in delays by monitoring system performance, such as throughput.

\subsubsection{Redundant CSI-RS Elimination}
By predicting the current CSI, the corresponding channel estimation process and CSI-RS overhead can be avoided, leading to increased data transmission rates and throughput. However, without direct estimation, evaluating the accuracy of the predicted CSI becomes a challenging issue. Additionally, the fixed observation and prediction window might not adapt to changing wireless channels, which can degrade prediction performance. Thus, the predictor should be able to adjust the observation and prediction windows based on changes in wireless channels, such as extending the observation window and shortening the prediction window for faster channel changes.

\subsubsection{Feedback Overhead Reduction}
For gNB-side prediction methods, if the current CSI is obtained through prediction, the UE does not need to feed back the corresponding CSI, thus reducing the feedback overhead compared with UE-side prediction methods. With the increasing number of antennas, CSI feedback can consume excessive bandwidth. Therefore, reducing feedback overhead without sacrificing accuracy has become an essential goal for channel prediction. Moreover, the impact of the CSI feedback method should be considered, as increasing the CSI compression ratio (CR) reduces prediction performance but saves more feedback overhead, while decreasing the CR does the opposite.
\subsection{Prediction Accuracy Evaluation}

Prediction accuracy-related KPIs can be divided into intermediate and eventual KPIs. Intermediate KPIs directly evaluate the accuracy of the predicted CSI, where NMSE and SGCS are commonly used to assess the prediction accuracy of the complete channel matrix and eigenvector, respectively. The performance of no prediction and non-AI prediction methods is usually used as a baseline for comparison. For gNB-side prediction, CSI compressed by the eType II codebook with no prediction can also serve as a baseline. Moreover, the eType II codebook with Doppler-domain compression is supported in Rel-19, providing an additional baseline.

Eventual KPIs, on the other hand, mainly use throughput to evaluate the effects of CSI prediction on systems, with throughput using ideal CSI as a performance upper bound. Intermediate KPIs are commonly adopted in existing CSI prediction works, but eventual KPIs are better suited to assess the effects of CSI prediction on the entire system, providing more accurate conclusions on the benefits of AI. For system-level simulation, both intermediate and eventual KPIs can be employed, making it more preferred over link-level simulation, which is limited to intermediate KPIs.

In terms of datasets, many current works rely on datasets based on statistical models from TR 38.901 \cite{901} for CSI prediction evaluation. Nevertheless, using real-world datasets obtained from field tests is essential for improving the AI model's realism. Unfortunately, collecting and labeling real-world data is laborious, presenting evaluation difficulties. Therefore, more efficient methods for generating datasets with less overhead should be developed, such as using digital twins to replicate real-world communication systems virtually \cite{imt,8424832,9899718,9897088}, which are closer to reality than current 3GPP dataset generation methods.

\subsection{Generalization Performance Evaluation}

The performance of NNs heavily depends on the datasets used for training and testing. If there is a mismatch between the features of these datasets, performance loss is inevitable. Hence, evaluating and improving the AI model's generalization performance to different CSI features is essential.

There are mainly three types of methods to enhance the generalization performance of AI-based prediction frameworks:

\begin{enumerate}
\item {\bf Generalized networks}: Training a generalized network with datasets from all possible scenarios and configurations is a prevailing approach for generalization improvement. However, updating the training datasets and retraining the network is necessary when new scenarios emerge, which increases dataset delivery and network training burden. Moreover, achieving satisfactory performance across diverse scenarios and configurations often necessitates a generalized network with high complexity, further escalating its operational overhead.

\begin{figure}[t]
  \centering
  \includegraphics[width=0.40\textwidth]{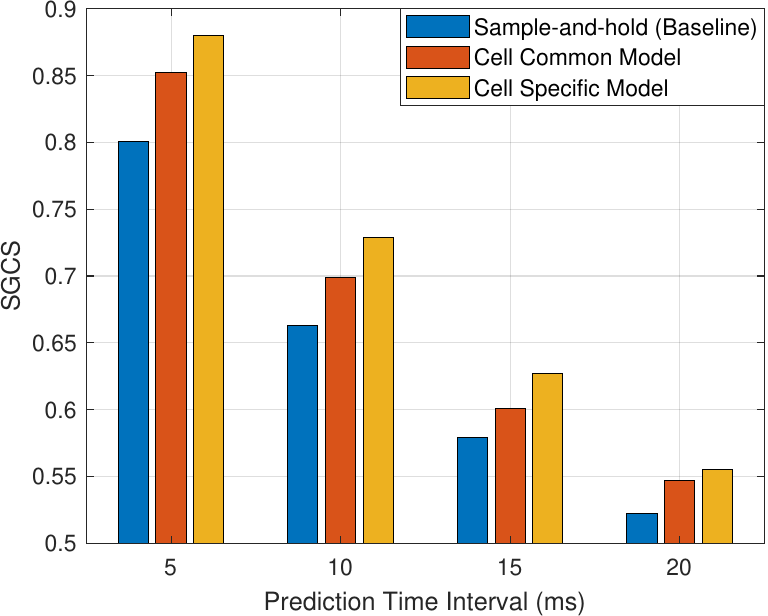}
  \caption{Comparison of cell specific, cell common, and no prediction methods \cite{cell}. }
  \label{cell}
  \vspace{-2mm}
\end{figure}
\item {\bf Specific networks}: In this method, multiple parallel networks are trained separately using CSI samples from different scenarios and configurations, and a classifier determines which network to use. Despite the deployment of multiple networks, the complexity of each network can be significantly reduced compared to the generalized network, enabling better performance with reduced overhead \cite{sum116}. Fig. \ref{cell} presents a comparative analysis of the performance between the cell-specific and cell-common prediction methods. The cell-common model is trained on a dataset collected from 19 cells, whereas the cell-specific model is trained specifically on a dataset from a single cell. This comparison is conducted within the context of an Urban Macrocell scenario with the same configurations described in Section 2.3.1.
It can be observed that the SGCS gain for the cell-common model has been improved by 2\% to 5\% compared to the cell-common model.

\item {\bf Transfer learning}: This method combines the former two. Specifically, one or a few generalized networks are pre-trained offline, and the UE fine-tunes the network with real-time collected CSI samples through online learning \cite{9175003,10381825,9442844}. It achieves better performance than generalized networks with less complexity and overhead. This method has been acknowledged and verified through evaluations conducted by the 3GPP \cite{2400795}. 

Moreover, we have conducted a simulation to assess the effectiveness of transfer learning. We investigated three cases where the predictor was trained and tested in different configurations. In the first case, the predictor was trained and tested solely under the UMi\_NLOS scenario. In the second case, it was trained under the UMa\_NLOS scenario but evaluated under the UMi\_NLOS scenario. Lastly, the predictor was initially trained under the UMa\_NLOS scenario and then fine-tuned with a limited UMi\_NLOS dataset before being tested under the UMi\_NLOS scenario. Additionally, we conducted our simulation at three different speeds: 10 km/h, 30 km/h, and 60 km/h, to indicate the impact of varying mobility conditions on the predictor's performance. 
\begin{figure}[t]
  \centering
  \includegraphics[width=0.40\textwidth]{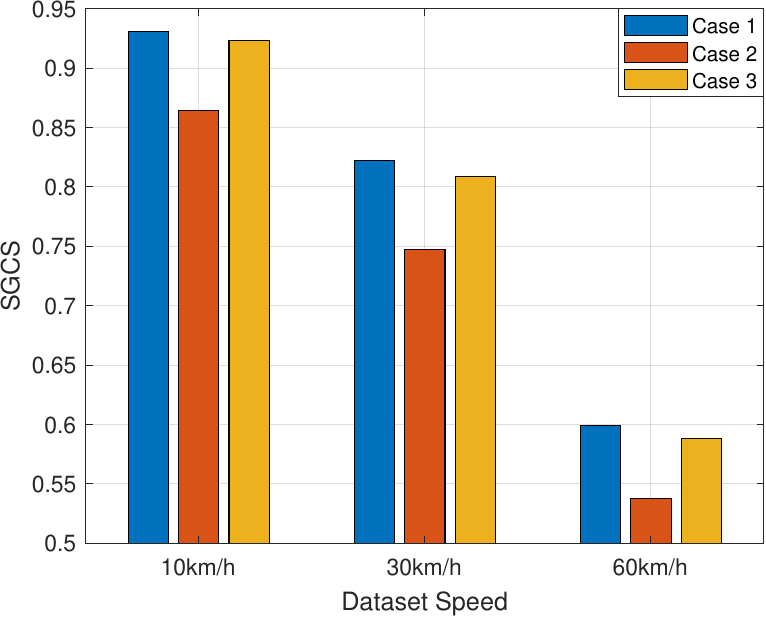}
  \caption{Comparison of different generalization cases. }
  \label{transfer}
  \vspace{-2mm}
\end{figure}

The simulation results are presented in Figure \ref{transfer}. It can be seen that the predictor's performance declines when confronted with scenarios that differ from its training dataset. However, transfer learning mitigates this performance degradation by fine-tuning with a limited dataset. This evaluation underscores the effectiveness of transfer learning as a method for enhancing generalization performance.

\end{enumerate}

\begin{figure}[t]
  \centering
  \includegraphics[width=0.40\textwidth]{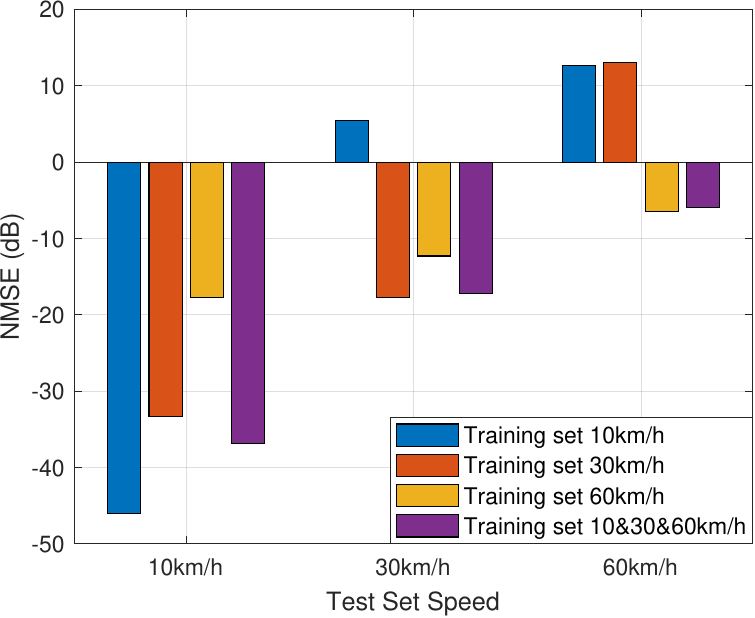}
  \caption{Comparison of NMSE for test sets and training sets with different speeds. \cite{speed}}
  \label{speed}
\end{figure}

To ensure the generalization performance of the CSI prediction frameworks, the following aspects should be emphasized:
\begin{enumerate}
\item {\bf Different scenarios}: The features of CSI highly depends on the propagation environment, and a UE may experience different scenarios due to its mobility. Hence, the framework's generalization performance across different scenarios is crucial. Methods like domain generalization \cite{9994050}, domain adaptation \cite{10411052}, and multi-task learning \cite{10262359} could be considered.

\item {\bf Different speeds}: CSI prediction utilizes the time correlation among CSI, which depends on the UE's speed and causes challenges in model switching. Fig. \ref{speed} \cite{speed} presents the generalization capabilities of AI-based CSI prediction across various UE speeds. 
Five historical raw channel matrices serve as inputs to a predictor, utilizing the multilayer perceptron-mixer architecture as its backbone, for the purpose of forecasting the raw channel matrix of a future instance.
It is evident that when the model is tested using a dataset with a UE speed differing from its training dataset, a significant degradation in performance is observed. However, when the model is trained on a mixed dataset containing diverse UE speeds, it exhibits good generalization performance across various speeds.
Therefore, utilizing a mixed dataset for training can be considered an effective approach to enhance the framework's generalization capabilities across different UE speeds.

\item {\bf Different SNRs}: Noise is inherent in the collected CSI, and the framework's generalization performance across different SNRs should be considered. Methods such as training the framework with noisy CSI or using a denoising module \cite{9076084,8896030} before training can be employed.
\end{enumerate}
\subsection{Complexity Evaluation}
The Rel-19 discussions on CSI prediction underscore the crucial need to assess the complexity increase associated with AI-based methods compared to non-AI approaches \cite{sum116}. For instance, the computational complexity of the AI-based prediction method depicted in Fig. \ref{non} is about 13 times that of the non-AI method \cite{oppo}.
Additionally, the model management and parameter update-related complexity should be considered for fair comparisons between non-AI and AI-based methods.

\begin{figure}[t]
  \centering
  \includegraphics[width=0.41\textwidth]{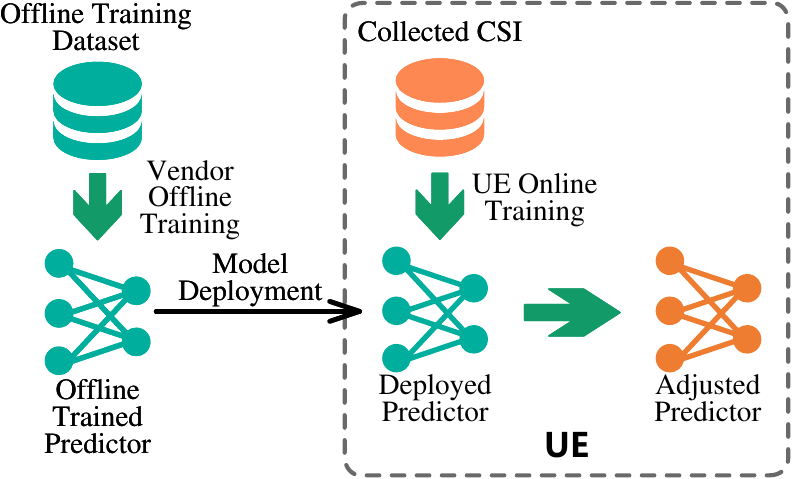}
  \caption{A UE-side online training method to train the UE-side model.}
  \label{trainue}
\end{figure}

Although some discussions suggest that with the continuous development of UE hardware, complexity should not be a hindrance to the deployment of AI techniques, 
it is still necessary to evaluate and reduce the UE-side model complexity to accommodate the UE's limited computation capacity, storage space, and power consumption in 5G-Advanced. The computational complexity can be measured using the metric of floating-point operations (FLOPs), while the model size can be measured using the memory storage in terms of NN parameters.
To this end, limiting the model size during NN design and employing NN pruning \cite{10279462}, knowledge distillation \cite{kd1} and weight quantization \cite{9136588} are effective approaches, and a trade-off between prediction accuracy and model complexity should be assessed. Meanwhile, it's worth noting that complexity and inference latency are hardware-dependent. Therefore, it's essential to report complexity-related KPIs alongside UE capabilities. These KPIs and UE capabilities can serve as the basis for model selection, for which effective selection algorithms need to be developed.

\subsection{Model Training}

\begin{figure*}[t]
  \centering
  \includegraphics[width=0.75\textwidth]{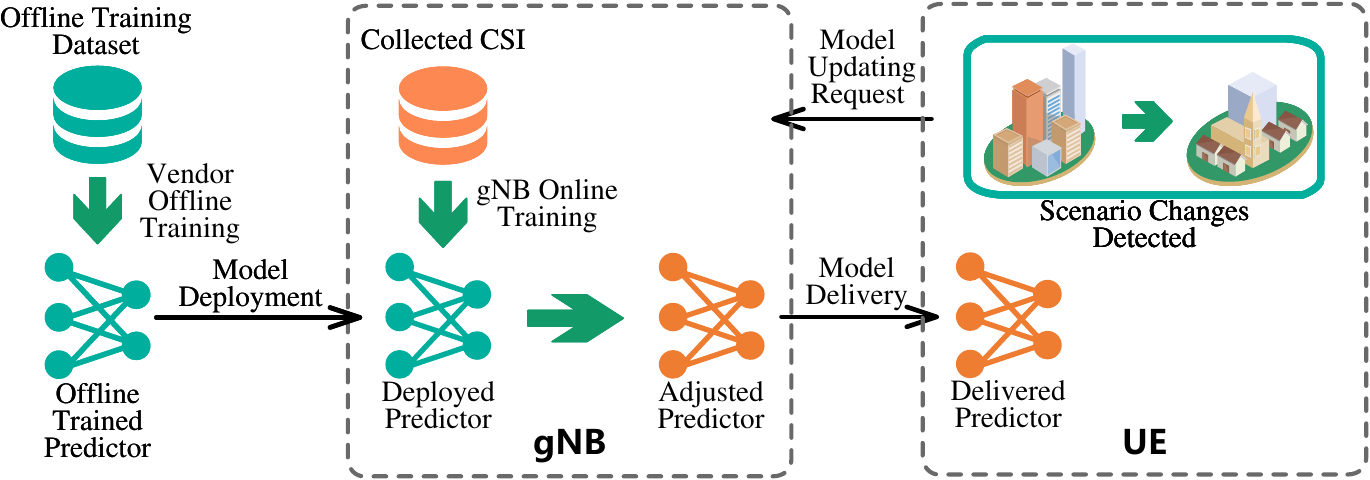}
  \caption{A gNB-side training method with model delivery to train the UE-side model, which performs online training at the gNB side and then delivers the trained model to the UE.}
  \label{traingnb}

\end{figure*}
\begin{figure}[t]
  \centering
  \includegraphics[width=0.41\textwidth]{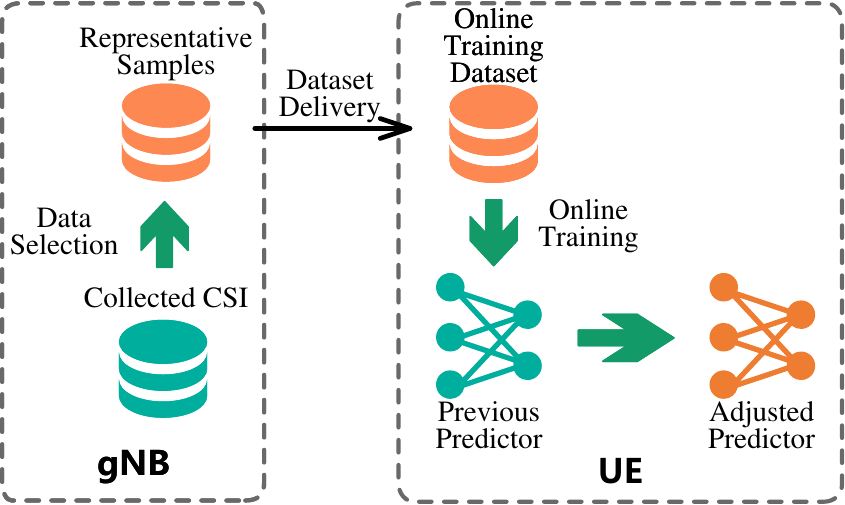}
  \caption{A dataset delivery-based training method, where the gNB collects sufficient CSI samples and delivers some representative samples to the UE.}
  \label{data}
  \vspace{-2mm}
\end{figure}
Most existing studies adopt offline training for evaluation, which allows for the use of more data and time for training and testing. However, relying solely on offline training from vendors may limit the framework's ability to generalize to potential scenarios, and updating the framework can be costly. Therefore, online training is necessary to achieve better performance in the practical deployment of AI-based frameworks.

However, online training for the UE and gNB poses unique challenges in comparison to offline training, particularly for the UE. The entity must first engage in an iterative process of collecting sufficient training data, which will be detailed in Section \ref{datacollection}, followed by online model training. Subsequently, the updated model must undergo validation by comparing its prediction output against real-time ground-truth data to ensure accuracy and reliability. This cyclic process necessitates advanced capabilities and expertise from the entity. 
As a result, the gNB is better suited for online training due to its robust capacity in computation, storage, power consumption, and communication. 

As for UE-side prediction, apart from performing online training on the UE side, as shown in Fig. \ref{trainue}, an alternative is to train the framework at the gNB side and then deliver the trained model to the UE, as shown in Fig. \ref{traingnb}. 
During this process, the gNB collects data transmitted from the UEs within its coverage area, often leveraging high-precision mechanisms such as the eType II codebook for optimal information gathering. Whenever the UE detects a change in its operating scenario, it initiates an updating request to the gNB. In response, the gNB then transmits the updated model, which has been trained using the aggregated cell data, to the requesting UE.  
Therefore, this approach proves beneficial for developing scenario-specific AI models \cite{2402267} and allows for a more centralized and efficient management of the prediction models, ensuring that UE can adapt to varying conditions in a timely and accurate manner. 
However, it is important to note that many companies consider the NN architecture as valuable intellectual property, which may pose challenges to the model delivery operation. Furthermore, the design of the NN heavily relies on the UE's capabilities, and compatibility issues may arise if the NN is designed without knowledge of the UE. Therefore, a collaborative approach regarding the UE's capabilities becomes necessary.

\subsection{Model Monitoring}

Model monitoring, recognized as an essential performance assurance for AI models, is considered a separate procedure within model management by the 3GPP. The process involves observing the AI model based on specific metrics and executing model management operations according to the monitoring results. Therefore, it is crucial to clearly define the monitoring metrics and their related operations. 

A potential performance monitoring solution for UE-side prediction involves implementing low-frequency estimation during the prediction phase to assess the accuracy of the predicted CSI \cite{sum116}. Specifically, the UE undertakes an estimation process for every few prediction intervals and subsequently calculates the accuracy of the predicted CSI against the estimated CSI. This process allows for prediction to proceed into subsequent intervals solely if the achieved accuracy surpasses a predetermined threshold.
For gNB-side prediction, extra feedback of this estimated CSI to the gNB is necessary to monitor the accuracy of the gNB-side predicted CSI. 
Another method proposed in \cite{pse} implements model monitoring by calculating the power spectral entropy (PSE) difference of the model input. When a significant PSE difference is detected, signifying potential environmental alterations, the UE performs an CSI estimation to assess the predicted CSI's accuracy.

Performance KPIs serve as common monitoring metrics. Through performance monitoring, the entity can make adjustments to the prediction framework's settings, such as the prediction window length, to adapt to performance fluctuations. Monitoring the performance of both the active model and inactive models is necessary. If the performance of an activated model falls below that of an inactive model, model switching occurs, where the current model is deactivated and replaced with the more effective one. Additionally, non-AI methods' performance should be monitored as well. If all available AI models perform poorly, the system should be able to fallback to non-AI methods. Collaborative communication between the UE and gNB is essential to ensure both sides are aware of each other's model operations. Misunderstandings can arise if one side experiences performance fluctuations due to model changes without knowledge of the other side's actions.

Data distribution is another monitoring metric. The NN model is designed to fit the distribution of the training dataset. Therefore, significant performance degradation can be anticipated if input samples do not align with the current model's data distribution. By monitoring data distribution, the entity can determine the most suitable model currently in use. In the event of substantial input drift, a model update may be required to maintain performance. Statistical methods, such as comparing probability distributions between the reference and monitored distributions, can be employed for evaluation. Metrics like entropy and Jensen-Shannon divergence \cite{10060176} can be useful in this context. Additionally, given the dynamic nature of wireless channels, the data distribution may experience sharp fluctuations. Methods like time domain averaging can be developed to enhance the robustness of model monitoring.

\subsection{Data Collection}
\label{datacollection}

Data collection is fundamental for model management, serving various purposes such as model training, inference, and monitoring. Different objectives require distinct data attributes, such as data size, latency, and validity, which should be considered separately. 
For model inference and monitoring, real-time data is essential to reflect the current system state. 
Therefore, data should be collected at short intervals with minimal delay. Conversely, model training requires a large and accurate dataset with less focus on latency. As such, the Rel-19 workshop proposes specifying a new data collection framework that supports the requirements for different purposes \cite{att}.
Model inference and monitoring require relatively fewer samples, making the overhead of timely data collection negligible. Though model training may require a larger dataset, infrequent collection is acceptable for such purposes \cite{2402026}. 

For UE-side prediction, in addition to collecting data directly from the UE, leveraging the gNB's data collection capabilities allows it to gather sufficient samples and deliver representative subsets to the UE, showcasing the main environmental features (see Fig. \ref{data}). This approach particularly benefits scenario-specific models.
The gNB can collect data via newly defined data collection RS. Subsequently, the representative training dataset can be transmitted by physical downlink shared channel or newly devised data (or control) channel. In Rel-19, it is anticipated that optimized data collection procedures will be established, potentially incorporating RS configuration requests, CSI measurement reporting mechanisms, and assistance signaling.
Moreover, managing vast amounts of data poses challenges to storage capacities. Consequently, data quantization becomes a viable strategy to mitigate storage costs, necessitating a balance between data quantity, quantization level, and model performance.
Furthermore, data quality is paramount to the framework's efficiency. Accurate CSI measurement techniques are imperative, and robust metrics (e.g., SNR and probability distribution) for data evaluation are essential.

\section{AI for CSI Prediction in Future Wireless Communication Systems}
\label{s4}

In this section, we explore the future of AI for CSI prediction for upcoming wireless systems. 
\subsection{Integrated Designs With CSI Feedback}

AI-based CSI feedback methods are also being explored in 5G-Advanced \cite{guo2022ai}. Compared to the current codebook-based CSI feedback, AI-based feedback methods offer improved accuracy with lower overhead. Moreover, the system performance gains of CSI prediction are influenced by the CSI feedback method. Therefore, effectively integrating the CSI prediction with the CSI feedback framework can result in superior performance. Furthermore, joint CSI compression and prediction has been identified as an objective for further study, and an accuracy gain of around 6\%-10\% has been observed over the Rel-18 eType II codebook.

For gNB-side prediction methods, CSI feedback becomes unnecessary when the predicted CSI closely matches the ideal CSI. Therefore, a performance threshold can be set during training and deployment to determine whether to use prediction or feedback. Prediction can continue if the performance remains above the threshold, and the system can switch to feedback when performance drops below the threshold. Additionally, for UE-side predictors with a prediction length over one, there exists a time correlation among the predicted CSI matrices, which can be leveraged to compress the predicted CSI using temporal-spatial-frequency domain compression.

During practical deployment, it is crucial to adjust the settings of the prediction and feedback frameworks based on performance changes. This calls for a clear understanding of how the prediction and feedback frameworks impact system performance. Moreover, if the traditional codebook-based feedback method is adopted, the CSI feedback mechanism in the temporal air interface can be retained. As changing the current feedback mechanism within a short period may be challenging, exploring CSI prediction with codebook-based feedback is worth investigating.

\subsection{Multi-Tasking Integration}
As highlighted in the Introduction, AI/ML is currently being explored in conjunction with three use cases in 5G-Advanced: enhanced CSI feedback, beam management, and improved positioning accuracy. Although their impacts are presently being studied and evaluated separately, future air interfaces are anticipated to integrate the CSI prediction framework into other use cases due to their high degree of interrelation. One particular task involves enhancing the positioning accuracy of the UE by considering the correlation between CSI and the UE's position. Through the integration of CSI prediction and positioning, these tasks can mutually benefit each other.

Furthermore, most of the current CSI prediction frameworks focus solely on improving prediction accuracy, aiming to obtain all the information of the predicted CSI. However, the optimization objectives for specific tasks have physical meanings that require only partial channel information. Therefore, evaluating prediction performance using task-related KPIs may be more relevant than prediction accuracy. Additionally, prediction accuracy can be linked to the supportable tasks. For instance, when prediction accuracy is low, only tasks not demanding high CSI accuracy, such as positioning or beamforming index, can be supported. As prediction accuracy improves, the scope of supported tasks can be broadened.

\subsection{Predictions for High-Speed Situations}
Current wireless communication systems face significant challenges in providing reliable communication in high-speed scenarios. Specifically, the ITU's recommendation for the vision of IMT-2030 emphasizes the necessity to enhance mobility, setting research targets of up to 500-1,000 km/h \cite{imt}. This objective encompasses advanced high-speed railway and airplane communication scenarios. Consequently, channel prediction in high-speed scenarios becomes a critical issue for the advancement of future wireless communication systems. 

In high-speed railway scenarios, the rapid movement of trains leads to low correlation between adjacent CSI, posing challenges for CSI prediction. Nevertheless, the train follows a fixed trajectory, and the environment around the base station generally remains relatively stable. This stability results in CSI changes that follow specific patterns, which can be utilized for CSI prediction. Prediction based on multimodal fusion \cite{10345638} is worth considering, where the prediction framework utilizes diverse information such as position and speed for accurate prediction. Moreover, the concept of integrated sensing and communication has been identified as a usage scenario for IMT-2030 \cite{imt}. This integration can provide high-precision position and speed information to enhance prediction accuracy \cite{9376324}. In airplane communication scenarios, although the flight paths of airplanes might not be consistent, the communication channels are sparse and primarily line-of-sight, allowing current prediction methods to achieve commendable performance.

\section{Conclusions}
\label{s5}
This article furnishes valuable insights into the application of AI for CSI prediction in 5G-Advanced and future wireless communication systems. It accentuates the benefits of using AI approaches in this context by juxtaposing non-AI and AI-based prediction methods. The discussions on standardization aspects and open problems illuminate key considerations for successful implementation, including the scope, performance evaluation, and model management, as well as prospects in future wireless communication systems. Nevertheless, to fully harness AI's potential in wireless communication systems, continued discussions and regulatory initiatives are essential. Future research should build upon the existing achievements and consensuses to expedite the deployment of AI for CSI prediction in 5G-Advanced and beyond.

\section*{ACKNOWLEDGEMENT}
This work was supported in part by the National Natural Science Foundation of China (NSFC) under Grant 62261160576, the Key Technologies R\&D Program of Jiangsu (Prospective and Key Technologies for Industry) under Grants BE2023022-1 and BE2023022, and the Fundamental Research Funds for the Central Universities under Grant 2242023K5003. The work was also supported in part by the National Natural Science Foundation of China (NSFC) under Grant 62401640 and in part by Guangdong Basic and Applied Basic Research Foundation under Grant 2023A1515110732. This work of Jun Zhang was supported partly by the Hong Kong Research Grants Council under the NSFC/RGC Collaborative Research Scheme grant CRS\_HKUST603/22.

\bibliographystyle{gbt7714-numerical}

\bibliography{main}

\begin{thebibliography}{62}
\providecommand{\natexlab}[1]{#1}
\providecommand{\url}[1]{#1}
\expandafter\ifx\csname urlstyle\endcsname\relax\else
  \urlstyle{same}\fi
\expandafter\ifx\csname href\endcsname\relax
  \DeclareUrlCommand\doi{\urlstyle{rm}}
  \def\eprint#1#2{#2}
\else
  \def\doi#1{\href{https://doi.org/#1}{\nolinkurl{#1}}}
  \let\eprint\href
\fi

\bibitem[Wang et~al.(Nov. 2017)Wang, Wen, Wang, Gao, Jiang, and Jin]{8233654}
WANG T, WEN C~K, WANG H, et~al.
\newblock Deep learning for wireless physical layer: {O}pportunities and
  challenges\allowbreak[J].
\newblock China Commun., Nov. 2017, 14\allowbreak (11): 92-111.

\bibitem[Chen et~al.(June 2023)Chen, Lin, Lee, Toskala, Sun, Chiasserini, and
  Liu]{10158439}
CHEN W, LIN X, LEE J, et~al.
\newblock 5{G}-advanced toward {6G}: {P}ast, present, and future\allowbreak[J].
\newblock IEEE J. Select. Areas Commun., June 2023, 41\allowbreak (6):
  1592-1619.

\bibitem[Wang et~al.(2nd Quart. 2023)]{wang2023road}
WANG C~X, et~al.
\newblock On the road to {6G}: Visions, requirements, key technologies, and
  testbeds\allowbreak[J].
\newblock IEEE Commun. Surveys Tuts., 2nd Quart. 2023, 25\allowbreak (2):
  905-974.

\bibitem[{ITU-R WP5D}(June 2023)]{imt}
{ITU-R WP5D}.
\newblock {Draft new Recommendation ITU-R M.[IMT.FRAMEWORK FOR 2030 AND BEYOND]
  - Framework and overall objectives of the future development of IMT for 2030
  and beyond}\allowbreak[R/OL].
\newblock June 2023.
\newblock
  \url{https://www.itu.int/md/meetingdoc.asp?lang=en\&parent=R19-WP5D-230612-TD-0905}.

\bibitem[Jin et~al.(June 2023)Jin, Liu, Zhang, Zhang, Lee, Farag, Zhu,
  Onggosanusi, Shafi, and Tataria]{10121037}
JIN H, LIU K, ZHANG M, et~al.
\newblock Massive {MIMO} evolution toward {3GPP R}elease 18\allowbreak[J].
\newblock IEEE J. Select. Areas Commun., June 2023, 41\allowbreak (6):
  1635-1654.

\bibitem[{WEN} et~al.(Oct. 2018){WEN}, {SHIH}, and {JIN}]{8322184}
{WEN} C, {SHIH} W, {JIN} S.
\newblock Deep learning for massive {MIMO CSI} feedback\allowbreak[J].
\newblock IEEE Wireless Commun. Lett., Oct. 2018, 7\allowbreak (5): 748-751.

\bibitem[Lu et~al.(Jan. 2019)Lu, Xu, Shen, Zhu, and Wang]{lu2018mimo}
LU C, XU W, SHEN H, et~al.
\newblock {MIMO channel information feedback using deep recurrent
  network}\allowbreak[J].
\newblock IEEE Commun. Lett., Jan. 2019, 23\allowbreak (1): 188-191.

\bibitem[Xiao et~al.(2024)Xiao, Wang, Li, et~al.]{10495862}
XIAO H, WANG Z, LI D, et~al.
\newblock {AI} enlightens wireless communication: {A} transformer backbone for
  {CSI} feedback\allowbreak[J].
\newblock China Commun., 2024.

\bibitem[Zhou et~al.(Jan. 2019)Zhou, Fang, Wang, Long, He, and Han]{8542687}
ZHOU P, FANG X, WANG X, et~al.
\newblock Deep learning-based beam management and interference coordination in
  dense mmwave networks\allowbreak[J].
\newblock IEEE Trans. Veh. Technol., Jan. 2019, 68\allowbreak (1): 592-603.

\bibitem[Liu et~al.(Feb. 2020)Liu, Wang, Boudreau, Sediq, and
  Abou-zeid]{8968715}
LIU Y, WANG X, BOUDREAU G, et~al.
\newblock Deep learning based hotspot prediction and beam management for
  adaptive virtual small cell in {5G} networks\allowbreak[J].
\newblock IEEE Trans. Emerg. Topics Comput. Intell., Feb. 2020, 4\allowbreak
  (1): 83-94.

\bibitem[Yang et~al.(Aug. 2021)Yang, Jin, Wen, Guo, Matthaiou, and
  Gao]{9390409}
YANG J, JIN S, WEN C~K, et~al.
\newblock Model-based learning network for 3-{D} localization in mmwave
  communications\allowbreak[J].
\newblock IEEE Trans. Wireless Commun., Aug. 2021, 20\allowbreak (8):
  5449-5466.

\bibitem[Gu et~al.(Dec. 2023)Gu, Yang, Gui, and Gacanin]{10163845}
GU H, YANG J, GUI G, et~al.
\newblock Triplet matchnet based indoor position method using {CSI} fingerprint
  similarity comparison\allowbreak[J].
\newblock IEEE Trans. Veh. Technol., Dec. 2023, 72\allowbreak (12):
  16905-16910.

\bibitem[{Moderator (Qualcomm)}(Dec. 2021)]{213599}
{Moderator (Qualcomm)}.
\newblock New {SI}: {S}tudy on artificial intelligence ({AI})/{Machine
  Learning} ({ML}) for {NR} air interface\allowbreak[R/OL].
\newblock {3GPP RP-213599}, Dec. 2021.
\newblock \url{https://www.3gpp.org/ftp/tsg ran/TSG RAN/TSGR
  94e/Docs/RP-213599.zip}.

\bibitem[Guo et~al.(June 2024)Guo, Wen, Jin, and Li]{guo2022ai}
GUO J, WEN C~K, JIN S, et~al.
\newblock {AI for CSI} feedback enhancement in {5G-Advanced}\allowbreak[J].
\newblock IEEE Wireless Commun., June 2024, 31\allowbreak (3): 169-176.

\bibitem[Guo et~al.(Dec. 2022)Guo, Wen, Jin, and Li]{9931713}
GUO J, WEN C~K, JIN S, et~al.
\newblock Overview of deep learning-based {CSI} feedback in massive {MIMO}
  systems\allowbreak[J].
\newblock IEEE Trans. Commun., Dec. 2022, 70\allowbreak (12): 8017-8045.

\bibitem[Love et~al.(Oct. 2008)Love, Heath, et~al.]{love2008overview}
LOVE D~J, HEATH R~W, et~al.
\newblock An overview of limited feedback in wireless communication
  systems\allowbreak[J].
\newblock IEEE J. Sel. Areas Commun., Oct. 2008, 26\allowbreak (8): 1341-1365.

\bibitem[Papazafeiropoulos(Feb. 2017)]{7473866}
PAPAZAFEIROPOULOS A~K.
\newblock Impact of general channel aging conditions on the downlink
  performance of massive {MIMO}\allowbreak[J].
\newblock IEEE Trans. Veh. Technol, Feb. 2017, 66\allowbreak (2): 1428-1442.

\bibitem[Luo et~al.(Jan. 2020)Luo, Ji, Wang, Chen, and Li]{8395053}
LUO C, JI J, WANG Q, et~al.
\newblock Channel state information prediction for 5{G} wireless
  communications: {A} deep learning approach\allowbreak[J].
\newblock IEEE Trans. Netw. Sci. Eng., Jan. 2020, 7\allowbreak (1): 227-236.

\bibitem[Jiang et~al.(Sept. 2022)Jiang, Cui, Ng, and Dai]{pre3}
JIANG H, CUI M, NG D~W~K, et~al.
\newblock Accurate channel prediction based on transformer: Making mobility
  negligible\allowbreak[J].
\newblock IEEE J. Sel. Areas Commun., Sept. 2022, 40\allowbreak (9): 2717-2732.

\bibitem[Chu et~al.(Dec. 2022)Chu, Liu, Lau, Jiang, and Yang]{pre4}
CHU M, LIU A, LAU V~K, et~al.
\newblock Deep reinforcement learning based end-to-end multi-user channel
  prediction and beamforming\allowbreak[J].
\newblock IEEE Trans. Wireless Commun., Dec. 2022, 21\allowbreak (12):
  10271-10285.

\bibitem[Sun et~al.(Dec. 2023)Sun, Zhang, Gui, Zhao, Gacanin, and
  Sari]{10174691}
SUN J, ZHANG Y, GUI G, et~al.
\newblock Interacting federated and transfer learning-aided {CSI} prediction
  for intelligent cellular networks\allowbreak[J].
\newblock IEEE Trans. Veh. Technol., Dec. 2023, 72\allowbreak (12):
  15776-15787.

\bibitem[Yang et~al.(Dec. 2020)Yang, Gao, Zhong, Ai, and Alkhateeb]{9175003}
YANG Y, GAO F, ZHONG Z, et~al.
\newblock Deep transfer learning-based downlink channel prediction for {FDD}
  massive {MIMO} systems\allowbreak[J].
\newblock IEEE Trans. Commun., Dec. 2020, 68\allowbreak (12): 7485-7497.

\bibitem[Zhang et~al.(Aug. 2021)Zhang, Wu, Liu, Xia, Pan, and Liu]{9439942}
ZHANG Y, WU Y, LIU A, et~al.
\newblock Deep learning-based channel prediction for {LEO} satellite massive
  {MIMO} communication system\allowbreak[J].
\newblock IEEE Wireless Commun. Lett., Aug. 2021, 10\allowbreak (8): 1835-1839.

\bibitem[Zhou et~al.(Jul. 2022)Zhou, Zhang, Ai, Xue, and Liu]{hsr}
ZHOU T, ZHANG H, AI B, et~al.
\newblock Deep-learning-based spatial–temporal channel prediction for smart
  high-speed railway communication networks\allowbreak[J].
\newblock IEEE Trans. Wireless Commun., Jul. 2022, 21\allowbreak (7):
  5333-5345.

\bibitem[Yuan et~al.(May. 2020)Yuan, Ngo, and Matthaiou]{8979256}
YUAN J, NGO H~Q, MATTHAIOU M.
\newblock Machine learning-based channel prediction in massive {MIMO} with
  channel aging\allowbreak[J].
\newblock IEEE Trans. Wireless Commun., May. 2020, 19\allowbreak (5):
  2960-2973.

\bibitem[Liu et~al.(June 2024)Liu, Hu, Wang, Zhang, Xue, and
  Matthaiou]{10404045}
LIU G, HU Z, WANG L, et~al.
\newblock A hypernetwork based framework for non-stationary channel
  prediction\allowbreak[J].
\newblock IEEE Trans. Veh. Technol., June 2024, 73\allowbreak (6): 8338-8351.

\bibitem[Kadambar et~al.(2023)Kadambar, Chavva, Lim, Goyal, Singh, Kumar, and
  Bal]{10333396}
KADAMBAR S, CHAVVA A~K~R, LIM C, et~al.
\newblock {Smart-CSI}: Deep learning based low complexity {CSI} prediction for
  beyond-5{G} systems\allowbreak[C]//\allowbreak
2023 IEEE 98th Veh. Technol. Conf. (VTC2023-Fall).
\newblock 2023: 1-5.

\bibitem[Shehzad et~al.(Apr. 2022)Shehzad, Rose, Wesemann, and Assaad]{9691478}
SHEHZAD M~K, ROSE L, WESEMANN S, et~al.
\newblock {ML}-based massive {MIMO} channel prediction: Does it work on
  real-world data?\allowbreak[J].
\newblock IEEE Wireless Commun. Lett., Apr. 2022, 11\allowbreak (4): 811-815.

\bibitem[Liu et~al.(Apr. 2024)Liu, Ma, Wang, and Qiao]{82336541}
LIU H, MA L, WANG Z, et~al.
\newblock Channel prediction for under water acoustic communication: {A} review
  and performance evaluation of algorithms\allowbreak[J].
\newblock Remote Sens., Apr. 2024, 16\allowbreak (9): 1546.

\bibitem[Soszka(2022)]{82336542}
SOSZKA M.
\newblock Fading channel prediction for 5{G} and 6{G} mobile communication
  systems\allowbreak[J].
\newblock Int. J. Electron. Telecommun., 2022, 68\allowbreak (1): 153-160.

\bibitem[Konstantinov et~al.(2019)Konstantinov and Pestryakov]{82336543}
KONSTANTINOV A~S, PESTRYAKOV A~V.
\newblock Fading channel prediction for 5{G}\allowbreak[C]//\allowbreak
Proc. Syst. Signal Synchronization Gener. Process. Telecommun. (SYNCHROINFO).
\newblock 2019: 1-7.

\bibitem[{Moderator (Qualcomm)}(Dec. 2023)]{234039}
{Moderator (Qualcomm)}.
\newblock New {WID} on study on artificial intelligence ({AI})/machine learning
  ({ML}) for {NR} air interface\allowbreak[R/OL].
\newblock {3GPP RP-234039}, Dec. 2023.
\newblock
  \url{https://www.3gpp.org/ftp/tsg_ran/TSG_RAN/TSGR_102/Info_for_workplan/new_approved_WID_12/RAN1_5/RP-234039.zip}.

\bibitem[{LG Electronics}(Apr. 2024)]{sum116}
{LG Electronics}.
\newblock Summary\#4 of {CSI} prediction\allowbreak[R/OL].
\newblock {3GPP R1-2403484}, Apr. 2024.
\newblock
  \url{https://www.3gpp.org/ftp/tsg\_ran/WG1\_RL1/TSGR1\_116b/Docs/R1-2403484.zip}.

\bibitem[Rao et~al.(Apr. 2024)Rao, Luo, Luo, Yi, Lei, and Cao]{10345638}
RAO X, LUO Z, LUO Y, et~al.
\newblock {MFFALoc: CSI}-based multifeatures fusion adaptive device-free
  passive indoor fingerprinting localization\allowbreak[J].
\newblock IEEE Internet Things J., Apr. 2024, 11\allowbreak (8): 14100-14114.

\bibitem[Pin et~al.(2021)Pin, Danny, He, Li, Bayesteh, Chen, Zhu, and
  Tong]{9376324}
PIN T, DANNY K, HE J, et~al.
\newblock Integrated sensing and communication in 6{G}: Motivations, use cases,
  requirements, challenges and future directions\allowbreak[C]//\allowbreak
IEEE 1st Int. Online Symp. Joint Commun. Sens. (JC\&S).
\newblock 2021: 1-6.

\bibitem[Wang et~al.(Feb. 2024)Wang, Yang, Cui, Xie, and Sun]{9994050}
WANG D, YANG J, CUI W, et~al.
\newblock {AirFi}: Empowering {WiFi}-based passive human gesture recognition to
  unseen environment via domain generalization\allowbreak[J].
\newblock IEEE Trans. Mobile Comput., Feb. 2024, 23\allowbreak (2): 1156-1168.

\bibitem[Liang et~al.(Jan. 2024)Liang, Wu, Li, Chang, Chen, Peng, and
  Xu]{10411052}
LIANG Y, WU W, LI H, et~al.
\newblock {DCS-Gait: A} class-level domain adaptation approach for cross-scene
  and cross-state gait recognition using {Wi-Fi CSI}\allowbreak[J].
\newblock IEEE Trans. Inf. Forensics Security, Jan. 2024, 19: 2997-3007.

\bibitem[Ye et~al.(Aug. 2020)Ye, Gao, Qian, Wang, and Li]{9076084}
YE H, GAO F, QIAN J, et~al.
\newblock Deep learning-based denoise network for {CSI} feedback in {FDD}
  massive {MIMO} systems\allowbreak[J].
\newblock IEEE Commun. Lett., Aug. 2020, 24\allowbreak (8): 1742-1746.

\bibitem[Erak et~al.(2023)Erak and Abou-Zeid]{10279462}
ERAK O, ABOU-ZEID H.
\newblock Accelerating and compressing deep neural networks for massive {MIMO
  CSI} feedback\allowbreak[C]//\allowbreak
IEEE Int. Conf. Commun. (ICC).
\newblock 2023: 1029-1035.

\bibitem[{OPPO}(Apr. 2024)]{oppo}
{OPPO}.
\newblock {Additional study on AI/ML-based CSI prediction}\allowbreak[R/OL].
\newblock {3GPP R1-2402318}, Apr. 2024.
\newblock \url{https://www.3gpp.org/ftp/tsg\_ran/WG1\_RL1/
  TSGR1\_116b/Docs/R1-2402318.zip}.

\bibitem[Kim et~al.(Jan. 2021)Kim, Kim, Lee, Jang, Choi, and Choi]{9210016}
KIM H, KIM S, LEE H, et~al.
\newblock Massive {MIMO} channel prediction: Kalman filtering {V}s. machine
  learning\allowbreak[J].
\newblock IEEE Trans. Commun., Jan. 2021, 69\allowbreak (1): 518-528.

\bibitem[{ZTE}(Feb. 2024)]{zte}
{ZTE}.
\newblock Evaluation on {AI CSI} feedback enhancement\allowbreak[R/OL].
\newblock {3GPP R1-2300171}, Feb. 2024.
\newblock
  \url{https://www.3gpp.org/ftp/tsg\_ran/WG1\_RL1/TSGR1\_112/Docs/R1-2300171.zip}.

\bibitem[{OPPO}(Apr. 2024)]{cell}
{OPPO}.
\newblock {Additional study on AI/ML-based CSI prediction}\allowbreak[R/OL].
\newblock {3GPP R1-2402318}, Apr. 2024.
\newblock
  \url{https://www.3gpp.org/ftp/tsg\_ran/WG1\_RL1/TSGR1\_116b/Docs/R1-2402318.zip}.

\bibitem[Li et~al.(Dec. 2023)Li, Guo, Wen, Jin, Han, and Wang]{10262359}
LI X, GUO J, WEN C~K, et~al.
\newblock Multi-task learning-based {CSI} feedback design in multiple
  scenarios\allowbreak[J].
\newblock IEEE Trans. Commun., Dec. 2023, 71\allowbreak (12): 7039-7055.

\bibitem[{Nokia, Nokia Shanghai Bell}(Feb. 2024)]{2400795}
{Nokia, Nokia Shanghai Bell}.
\newblock {AI/ML} for {CSI} prediction\allowbreak[R/OL].
\newblock {3GPP R1-2400795}, Feb. 2024.
\newblock
  \url{https://www.3gpp.org/ftp/tsg\_ran/WG1\_RL1/TSGR1\_116b/Docs/R1-2400795.zip}.

\bibitem[Fujitsu(Apr. 2024)]{speed}
FUJITSU.
\newblock {Discussion on CSI prediction with AI/ML}\allowbreak[R/OL].
\newblock {3GPP R1-2402788}, Apr. 2024.
\newblock
  \url{https://www.3gpp.org/ftp/tsg\_ran/WG1\_RL1/TSGR1\_116b/Docs/R1-2402788.zip}.

\bibitem[{ZTE}(Apr. 2024)]{2402267}
{ZTE}.
\newblock Discussion on study for other aspects of {AI/ML} model and
  data\allowbreak[R/OL].
\newblock {3GPP R1-2402267}, Apr. 2024.
\newblock
  \url{https://www.3gpp.org/ftp/tsg\_ran/WG1\_RL1/TSGR1\_116b/Docs/R1-2402267.zip}.

\bibitem[{MediaTek Inc.}(Aug. 2023)]{pse}
{MediaTek Inc.}
\newblock Evaluation on {AI/ML for CSI} feedback enhancement\allowbreak[R/OL].
\newblock {3GPP R1-2308053}, Aug. 2023.
\newblock
  \url{https://www.3gpp.org/ftp/tsg\_ran/WG1\_RL1/TSGR1\_114/Docs/R1-2308053.zip}.

\bibitem[{{AT\&T}}(June 2023)]{att}
{{AT\&T}}.
\newblock {Overview of Rel-19 AI/ML for NR Air Interface and
  NG-RAN}\allowbreak[R/OL].
\newblock {3GPP RWS-230473}, June 2023.
\newblock \url{https://www.3gpp.org/ftp/tsg\_ran/TSG\_RAN/TSGR\_AHs/2023\_06\_
  RAN\_Rel-19\_WS/Docs/RWS-230473.zip}.

\bibitem[{Huawei}(Apr. 2024)]{2402026}
{Huawei}.
\newblock Discussion on {AI/ML for CSI} compression\allowbreak[R/OL].
\newblock {3GPP R1-2402026}, Apr. 2024.
\newblock
  \url{https://www.3gpp.org/ftp/tsg\_ran/WG1\_RL1/TSGR1\_116b/Docs/R1-2402026.zip}.

\bibitem[Hwang et~al.(1998, vol. 2)Hwang and Winters]{sos}
HWANG J~K, WINTERS J.
\newblock Sinusoidal modeling and prediction of fast fading
  processes\allowbreak[C]//\allowbreak
Proc. IEEE Global Commun. Conf. (GLOBECOM).
\newblock 1998, vol. 2: 892-897.

\bibitem[Chen et~al.(Feb. 2022)Chen, Guo, Wen, Jin, Li, and Yang]{chen2021deep}
CHEN M, GUO J, WEN C~K, et~al.
\newblock Deep learning-based implicit {CSI} feedback in massive
  {MIMO}\allowbreak[J].
\newblock IEEE Trans. Commun., Feb. 2022, 70\allowbreak (2): 935-950.

\bibitem[{3GPP}(Oct. 2019)]{901}
{3GPP}.
\newblock {Study on channel model for frequencies from 0.5 to 100
  GHz}\allowbreak[R].
\newblock Version 16.0.0, document TR 38.901, Oct. 2019.

\bibitem[El~Saddik(Apr. 2018)]{8424832}
EL~SADDIK A.
\newblock Digital twins: The convergence of multimedia
  technologies\allowbreak[J].
\newblock IEEE MultiMedia, Apr. 2018, 25\allowbreak (2): 87-92.

\bibitem[Mihai et~al.(4th Quart. 2022)Mihai, Yaqoob, Hung, Davis, Towakel,
  Raza, Karamanoglu, Barn, Shetve, Prasad, Venkataraman, Trestian, and
  Nguyen]{9899718}
MIHAI S, YAQOOB M, HUNG D~V, et~al.
\newblock Digital twins: A survey on enabling technologies, challenges, trends
  and future prospects\allowbreak[J].
\newblock IEEE Commun. Surveys Tuts., 4th Quart. 2022, 24\allowbreak (4):
  2255-2291.

\bibitem[Vilas~Boas et~al.(Sept. 2022)Vilas~Boas, Zirwas, and Haardt]{9897088}
VILAS~BOAS B, ZIRWAS W, HAARDT M.
\newblock Machine learning for {CSI} recreation in the digital twin based on
  prior knowledge\allowbreak[J].
\newblock IEEE Open J. Commun. Soc., Sept. 2022, 3: 1578-1591.

\bibitem[Zhang et~al.(Mar. 2024)Zhang, Wang, Lu, and Zhang]{10381825}
ZHANG X, WANG J, LU Z, et~al.
\newblock Continuous online learning-based {CSI} feedback in massive {MIMO}
  systems\allowbreak[J].
\newblock IEEE Commun. Lett., Mar. 2024, 28\allowbreak (3): 557-561.

\bibitem[Zeng et~al.(Dec. 2021)Zeng, Sun, Gui, Adebisi, Ohtsuki, Gacanin, and
  Sari]{9442844}
ZENG J, SUN J, GUI G, et~al.
\newblock Downlink {CSI} feedback algorithm with deep transfer learning for
  {FDD} massive {MIMO} systems\allowbreak[J].
\newblock IEEE Trans. Cognit. Commun. Netw., Dec. 2021, 7\allowbreak (4):
  1253-1265.

\bibitem[Jin et~al.(Jan. 2020)Jin, Zhang, Ai, and Zhang]{8896030}
JIN Y, ZHANG J, AI B, et~al.
\newblock Channel estimation for {mmWave} massive {MIMO} with convolutional
  blind denoising network\allowbreak[J].
\newblock IEEE Commun. Lett., Jan. 2020, 24\allowbreak (1): 95-98.

\bibitem[Hinton et~al.({2015, [online]. Available})Hinton, Vinyals, and
  Dean]{kd1}
HINTON G, VINYALS O, DEAN J.
\newblock Distilling the knowledge in a neural network\allowbreak[A].
\newblock 2015, [online]. Available: http://arxiv.org/abs/2108.02397.

\bibitem[Guo et~al.(Aug. 2020)Guo, Wang, Wen, Jin, and Li]{9136588}
GUO J, WANG J, WEN C~K, et~al.
\newblock Compression and acceleration of neural networks for
  communications\allowbreak[J].
\newblock IEEE Wireless Commun., Aug. 2020, 27\allowbreak (4): 110-117.

\bibitem[Inoue et~al.(2023)Inoue, Ohtsuki, Yamamoto, and Gui]{10060176}
INOUE M, OHTSUKI T, YAMAMOTO K, et~al.
\newblock Evaluation of source data selection for {DTL} based {CSI} feedback
  method in {FDD} massive {MIMO} systems\allowbreak[C]//\allowbreak
IEEE Consumer Commun. Netw. Conf. (CCNC).
\newblock 2023: 182-187.

\end{thebibliography}
\biographies
\begin{CCJNLbiography}{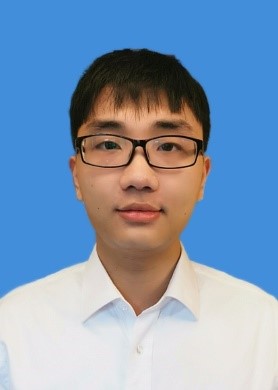}{Chengyong Jiang}
	received the B.S. degree from the Xidian University, Xi'an, China, in 2021. He is currently pursuing his Ph.D. degree in information and communications engineering with Southeast University, Nanjing, China. His current research interests include deep learning application in wireless communication, massive MIMO, and the standardization of AI in mobile communication systems.
\end{CCJNLbiography}

\begin{CCJNLbiography}{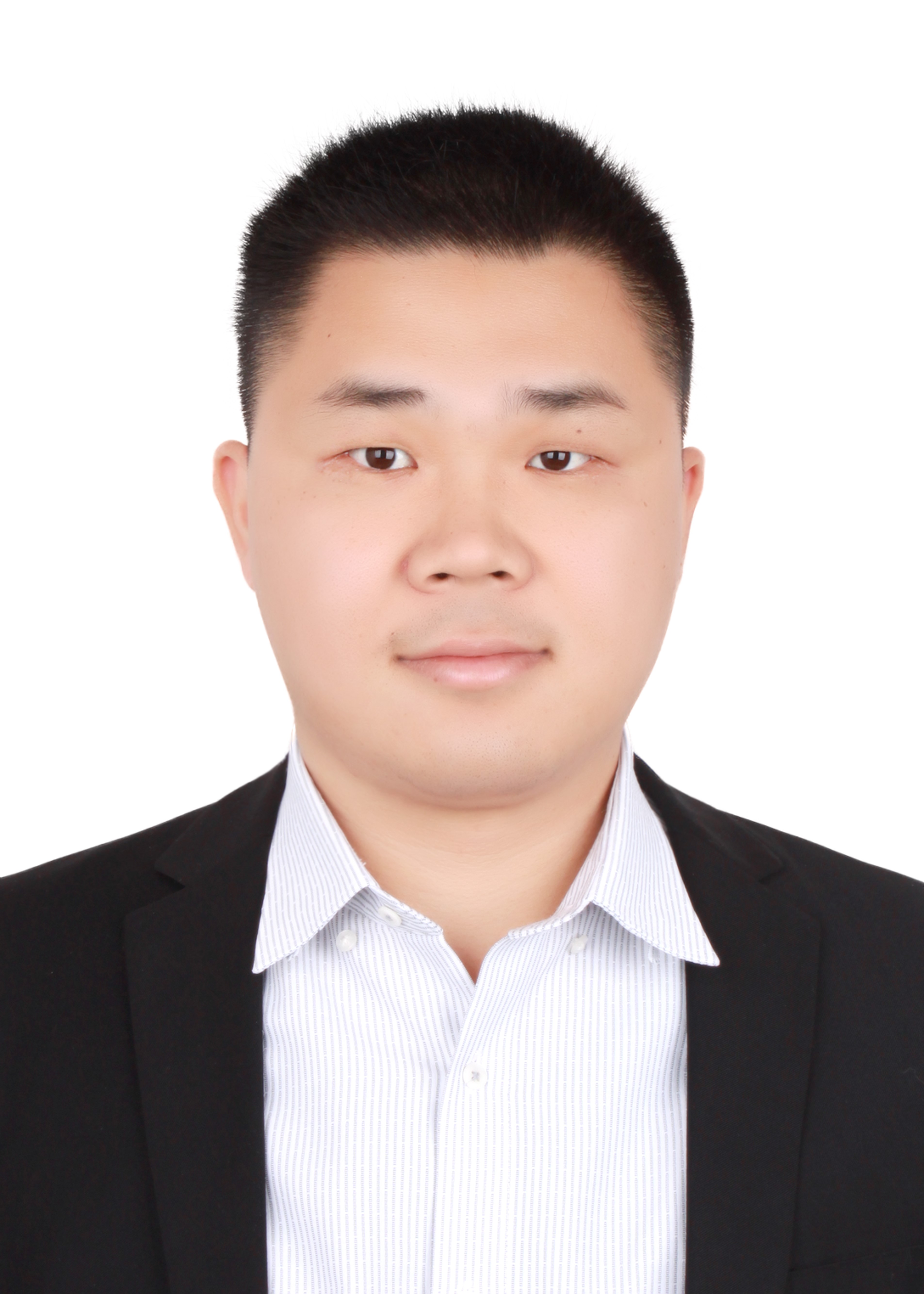}{Jiajia Guo}
	received the B.S. degree from the Nanjing University of Science and Technology, Nanjing, China, in 2016, the M.S. degree from the University of Science and Technology of China, Hefei, China, in 2019, and the Ph.D. degree in information and communications engineering from Southeast University, Nanjing, China, in 2023. Currently, he holds the position of Research Assistant Professor in the Department of Electronic and Computer Engineering (ECE) at The Hong Kong University of Science and Technology (HKUST). His research interests focus on AI-native air interfaces, massive MIMO, ISAC, and large AI models. His contributions were recognized as one of the Top 10 Science and Technology Advances in the Information and Communication field in China. Additionally, he received the 2023 First Prize of the Natural Science Award from the Chinese Institute of Electronics and the Best Doctoral Thesis Awards from both the Chinese Institute of Electronics and Jiangsu Province, China.

\end{CCJNLbiography}

\begin{CCJNLbiography}{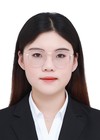}{Xiangyi Li}
	was born in Beijing, China, in 1994. She received her B.S. degree from the School of Mathematics, Tianjin University, Tianjin, China, in 2017, and her M.S. degree from the Centre for Applied Mathematics, Tianjin University, in 2020. She is currently working towards her Ph.D. degree in information and communications engineering, Southeast University, China. Her main research focuses on deep learning application in wireless communication and massive MIMO systems.
\end{CCJNLbiography}

\begin{CCJNLbiography}{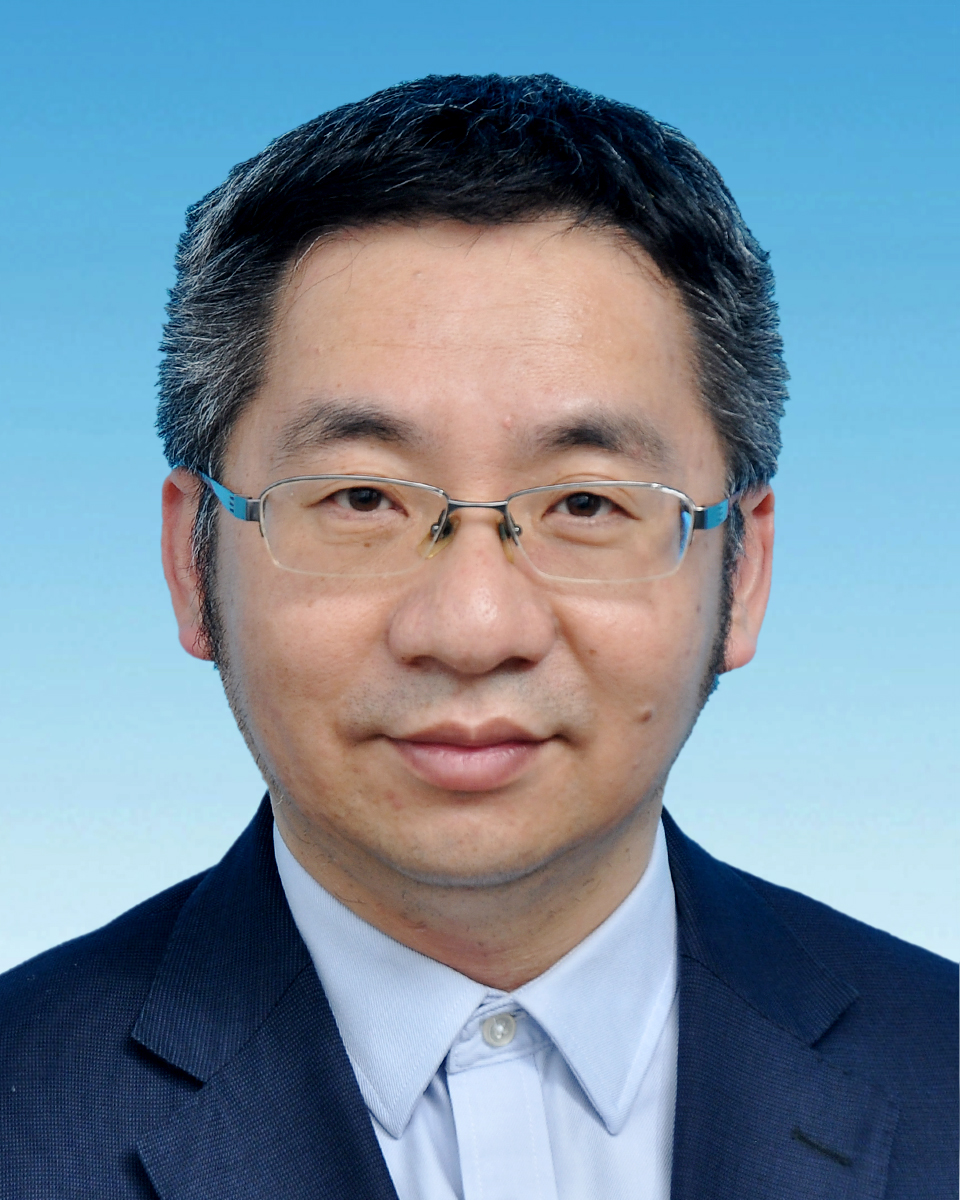}{Shi Jin}
	 received the B.S. degree in communications engineering from Guilin University of Electronic Technology, Guilin, China, in 1996, the M.S. degree from Nanjing University of Posts and Telecommunications, Nanjing, China, in 2003, and the Ph.D. degree in information and communications engineering from Southeast University, Nanjing, in 2007. From June 2007 to October 2009, he was a Research Fellow with the Adastral Park Research Campus, University College London, London, U.K. He is currently with the Faculty of the National Mobile Communications Research Laboratory, Southeast University. His research interests include wireless communications, random matrix theory, and information theory. He and his coauthors received the 2011 IEEE Communications Society Stephen O. Rice Prize Paper Award in the field of communication theory, the IEEE Vehicular Technology Society 2023 Jack Neubauer Memorial Award, the 2022 Best Paper Award, and the 2010 Young Author Best Paper Award by the IEEE Signal Processing Society. He is serving as an Area Editor for IEEE TRANSACTIONS ON COMMUNICATIONS and IET Electronics Letters. He was an Associate Editor of IEEE TRANSACTIONS ONWIRELESS COMMUNICATIONS, IEEE COMMUNICATIONS LETTERS,and IET Communications.
\end{CCJNLbiography}

\begin{CCJNLbiography}{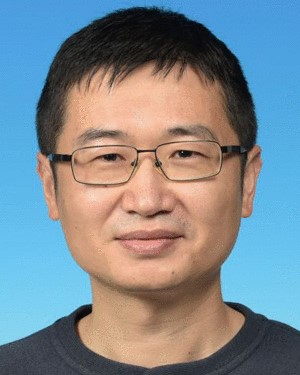}{Jun Zhang}
	received the B.Eng. degree in Electronic Engineering from the University of Science and Technology of China in 2004, the M.Phil. degree in Information Engineering from the Chinese University of Hong Kong in 2006, and the Ph.D. degree in Electrical and Computer Engineering from the University of Texas at Austin in 2009. He is an Associate Professor in the Department of Electronic and Computer Engineering at the Hong Kong University of Science and Technology. His research interests include wireless communications and networking, mobile edge computing and edge AI, and integrated AI and communications. Dr. Zhang co-authored the book Fundamentals of LTE (Prentice-Hall, 2010). He is a co-recipient of several best paper awards, including the 2021 Best Survey Paper Award of the IEEE Communications Society, the 2019 IEEE Communications Society \& Information Theory Society Joint Paper Award, and the 2016 Marconi Prize Paper Award in Wireless Communications. Two papers he co-authored received the Young Author Best Paper Award of the IEEE Signal Processing Society in 2016 and 2018, respectively. He also received the 2016 IEEE ComSoc Asia-Pacific Best Young Researcher Award. He is an Area Editor of IEEE Transactions on Wireless Communications and IEEE Transactions on Machine Learning in Communications and Networking. He is an IEEE Fellow and an IEEE ComSoc Distinguished Lecturer.

\end{CCJNLbiography}

\end{document}